\documentclass[journal, romanappendices]{IEEEtran}
\ifCLASSINFOpdf
\else
\fi
\usepackage[utf8]{inputenc} 
\usepackage[T1]{fontenc}
\usepackage{url}
\usepackage{ifthen}
\usepackage{cite}
\usepackage{graphicx}
\usepackage{hyperref}
\usepackage{amsmath}
\usepackage{amssymb}
\usepackage{bm}
\usepackage{bbm}
\usepackage{algorithmicx}
\usepackage{algorithm}
\usepackage{algpseudocode}
\usepackage{array}
\usepackage{booktabs}
\usepackage{multirow}
\usepackage{makecell}
\usepackage{mathtools}
\usepackage{mathrsfs}
\usepackage{xcolor}
\usepackage{balance}

\newtheorem{definition}{Definition}
\newtheorem{theorem}{Theorem}

\newtheorem{corollary}{Corollary}

\newcommand{\N}{\mathbb{N}}

\newcommand{\R}{\mathbb{R}}

\newcommand{\PP}{\mathsf{P}}
\newcommand{\X}{\mathcal{X}}
\newcommand{\Y}{\mathcal{Y}}
\newcommand{\U}{\mathcal{U}}

\newcommand{\G}{\mathcal{G}}
\newcommand{\E}{\mathbb{E}}

\newcommand{\Prob}{\mathbb{P}}

\newcommand*\diff{\mathop{}\!\mathrm{d}}

\newcommand{\Paren}[1]{\left(#1\right)}

\newcommand{\bmg}{\bm{g}}

\DeclareMathOperator{\st}{s. t.}

\DeclareMathOperator{\diag}{diag}

\allowdisplaybreaks

\begin{document}
%
\title{Systematic Transmission With Fountain Parity Checks for Erasure Channels With Stop Feedback}
%
%
%

\author{Hengjie~Yang,~\IEEEmembership{Member,~IEEE}
  and~Richard~D.~Wesel,~\IEEEmembership{Fellow,~IEEE}
\thanks{This work was supported by the National Science Foundation (NSF) under Grant CCF-1955660. Any opinions, findings, and conclusions or recommendations expressed in this material are those of the authors and do not necessarily reflect views of NSF.

Hengjie Yang is with Qualcomm Technologies, Inc., San Diego, CA 92121 USA (e-mail: hengjie.yang@ucla.edu).

Richard D. Wesel is with the Department of Electrical and Computer Engineering, University of California at Los Angeles, Los Angeles, CA 90095 USA (e-mail: wesel@ucla.edu).
}}

\maketitle

\begin{abstract}
In this paper, we present new achievability bounds on the maximal achievable rate of variable-length stop-feedback (VLSF) codes operating over a binary erasure channel (BEC) at a fixed message size $M = 2^k$. We provide new bounds for VLSF codes with zero error, infinite decoding times and with nonzero error, finite decoding times. Both new achievability bounds are proved by constructing a new VLSF code that employs systematic transmission of the first $k$ bits followed by random linear fountain parity bits decoded with a rank decoder. For VLSF codes with infinite decoding times, our new bound outperforms the state-of-the-art result for BEC by Devassy \emph{et al.} in 2016. We also give a negative answer to the open question Devassy \emph{et al.} put forward on whether the $23.4\%$ backoff to capacity at $k = 3$ is fundamental. For VLSF codes with finite decoding times, numerical evaluations show that the achievable rate for VLSF codes with a moderate number of decoding times closely approaches that for VLSF codes with infinite decoding times.
\end{abstract}


%
\IEEEpeerreviewmaketitle

\section{Introduction}
%
%
%
%
\IEEEPARstart{I}{n} a point-to-point communication system with stop feedback, the decoder decides on the fly when to stop transmission and sends a $1$-bit acknowledgement (ACK) or negative acknowledgement (NACK) symbol via the noiseless feedback channel informing the transmitter whether to stop or continue transmission. Meanwhile, the transmitter cannot utilize the stop-feedback symbol to design the next code symbol. Polyanskiy \emph{et al.} \cite{Polyanskiy2011} formalized this type of code as the \emph{variable-length stop-feedback (VLSF)} code. The VLSF code is of practical interest since it includes the hybrid automatic repeat request and incremental redundancy. Polyanskiy \emph{et al.} showed that even with such a limited use of feedback, the maximal achievable rate of VLSF code is significantly better than that of the fixed-length code in the nonasymptotic regime. Earlier works have also studied various aspects of VLSF codes, including the performance in error-exponent regime \cite{Forney1968}, performance for random linear codes over the binary erasure channel (BEC) \cite{Heidarzadeh2019}, and performance when noisy stop feedback is present \cite{Ostman2019}.

This paper focuses on the BEC with stop feedback and seeks a nonasymptotic achievability bound for the VLSF setup. To the best of our knowledge, Devassy \emph{et al.} obtained the state-of-the-art achievability \cite[Theorem 9]{Devassy2016} and converse bounds \cite[Corollary 6]{Devassy2016} for VLSF codes operating over a BEC. Note that constructing VLSF codes for the BEC is equivalent to constructing rateless erasure codes. Motivated by this observation, Devassy \emph{et al.} derived the achievability bound by analyzing a family of random linear fountain codes \cite[Chapter 50]{MacKay2005} of message size $2^k$ and by using a rank decoder. The rank decoder keeps track of the rank of the generator matrix associated with unerased received symbols. As soon as the rank equals $k$, the decoder stops transmission by sending an ACK symbol and reproduces the $k$-bit message with zero error using the inverse of the generator matrix. However, their achievability bound implies that the ratio of maximal achievable rate to capacity is only a function of message length $k$ and the ratio attains the maximum of $76.6\%$ at $k = 3$. Hence, they posed the question whether the $23.4\%$ backoff percentage to capacity at $k = 3$ is fundamental.

Unlike Devassy \emph{et al.}'s approach, we adopt a new coding scheme called \emph{systematic transmission followed by random linear fountain coding (ST-RLFC)}. Namely, the transmitter simply transmits the first $k$ message bits in the first $k$ time instants. After that, the transmitter employs a random linear fountain code to generate parity bits. Specifically, starting the $(k+1)$th time instant, both the encoder and decoder select the same nonzero base vector in $\{0, 1\}^k$ according to the common randomness. The encoder produces the code symbol by linearly combining the message bits using the selected base vector. The decoder is still the same rank decoder.

Our contributions in this paper are as follows.
\begin{itemize}
  \item By analyzing the ST-RLFC scheme, we present a new achievability bound for VLSF codes of message size $M = 2^k$ and zero error probability. Our new bound outperforms Devassy \emph{et al.}'s result \cite[Theorem 9]{Devassy2016}. In addition, the new bound implies that the $23.4\%$ backoff percentage to capacity reported by Devassy \emph{et al.} is \emph{not} fundamental. On the contrary, the new bound indicates that the backoff percentage is proportional to the erasure probability at any given message length $k$. If the erasure probability is zero, there is no backoff from capacity.
  \item The new VLSF achievability bound for zero-error VLSF codes implies a new achievability bound for VLSF codes constrained to have finite decoding times and a nonzero error probability. Numerical computations show that when number of decoding times $m$ is small, a slight increase in $m$ can dramatically improve the achievable rate. However, when $m$ is moderately large (for instance, $m = 16$ for erasure probability $0.5$), the achievable rate closely approaches that for $m = \infty$.
\end{itemize}

The remainder of this paper is organized as follows. Section \ref{sec: preliminary} introduces the notation and the VLSF code, and presents previously known bounds for VLSF codes operating over a BEC. Section \ref{sec: main results} provides the new VLSF achievability bound and its implications. Section \ref{sec: proofs} includes proofs of the main results. Section \ref{sec: conclusion} concludes the paper.

\section{Preliminaries}\label{sec: preliminary}

\subsection{Notation}

Let $\N = \{0, 1,\dots\}$, $\N_+ = \N\setminus\{0\}$, $\N_{\infty} = \N\cup\{\infty\}$ be the set of natural numbers, positive integers, and extended natural numbers, respectively. For $i\in\N$, $[i]\triangleq \{1,2,\dots, i\}$. We use $x_i^j$ to denote a sequence $(x_i, x_{i+1}, \dots, x_j)$, $1\le i\le j$. We denote by $\bm{e}_i\in\R^{k\times 1}$ the $k$-dimensional natural base vector with $1$ at index $i$ and $0$ everywhere else, $1\le i\le k$. We denote the distribution of a random variable $X$ by $\PP_X$.

\subsection{VLSF Codes}
We consider a BEC with input alphabet $\X = \{0, 1\}$, output alphabet $\Y = \{0, ?, 1\}$, and erasure probability $p\in[0, 1)$. A VLSF code for BEC with finite decoding times is defined as follows.
\begin{definition}\label{def: VLSF code}
An $(l, n_1^m, M, \epsilon)$ VLSF code, where $l > 0$, $m\in\N_{\infty}$, $n_1^m\in\N^m$ satisfying $n_1 < n_2 < \cdots < n_m$, $M\in\N_+$, and $\epsilon\in(0, 1)$, is defined by:
\begin{itemize}
  \item [1)] A finite alphabet $\U$ and a probability distribution $\PP_U$ on $\U$ defining the common randomness random variable $U$ that is revealed to both the transmitter and the receiver before the start of the transmission.
  \item [2)] A sequence of encoders $f_n: \U\times [M] \to \X$, $n = 1,2,\dots, n_m$, defining the channel inputs
    \begin{align}
      X_n = f_n(U, W),
    \end{align}
    where $W\in[M]$ is the equiprobable message.
  \item [3)] A non-negative integer-valued random stopping time $\tau\in\{n_1, n_2, \dots, n_m\}$ of the filtration generated by $\{U, Y^{n_i}\}_{i=1}^m$ that satisfies the average decoding time constraint
    \begin{align}
      \E[\tau] \le l.
    \end{align}
  \item [4)] $m$ decoding functions $g_{n_i}: \U\times \Y^{n_i}\to [M]$, providing the best estimate of $W$ at time $n_i$, $i\in[m]$. The final decision $\hat{W}$ is computed at time instant $\tau$, i.e., $\hat{W} = g_{\tau}(U, Y^\tau)$ and must satisfy the average error probability constraint
    \begin{align}
      P_e \triangleq \Prob[\hat{W}\ne W] \le \epsilon.
    \end{align}
\end{itemize}
\end{definition}
Comparing to Polyanskiy \emph{et al.}'s VLSF code definition \cite{Polyanskiy2011}, the primary distinctions are two-fold. First, the VLSF code is allowed to have finite decoding times rather than infinite decoding times. As a result, the stopping time is constrained within these decoding times. Second, both the expected blocklength and error probability constraints correspond to the given sequence of decoding times rather than $\N$.

In this paper, we focus on upper bounding the average blocklength $\E[\tau]$ of $(l, \N, 2^k, 0)$ VLSF code with $m = \infty$ and $(l, n_1^m, 2^k, \epsilon)$ VLSF code with $m < \infty$. The rate of an $(l, n_1^m, M, \epsilon)$ VLSF code is defined by 
\begin{align}
    R \triangleq \frac{\log M}{\E[\tau]}.
\end{align}

\subsection{Previous Results for VLSF Codes over BECs}

For the BEC, the decoder has the ability to identify the correct message whenever only a single codeword is compatible with the unerased channel outputs up to that point. By exploiting this fact and utilizing the RLFC, Devassy \emph{et al.} \cite{Devassy2016} obtained state-of-the-art achievability bound for zero-error VLSF codes with message size $M$ that is a power of $2$.
\begin{theorem}[Theorem 9, \cite{Devassy2016}]\label{theorem: BEC achievability}
  For each integer $k\ge 1$, there exists an $(l, \N, 2^k, 0)$ VLSF code for a BEC$(p)$ with
    \begin{align}
        l \le \frac{1}{C}\Paren{k + \sum_{i = 1}^{k-1} \frac{2^i - 1}{2^k - 2^i} }, \label{eq: Devassy bound}
    \end{align}
\end{theorem}

Note that the second term in parentheses of \eqref{eq: Devassy bound} is bounded by the Erd\"os-Borwein constant $c = 1.60669515...$ (OEIS: A065442),
\begin{align}
  \sum_{i = 1}^{k-1} \frac{2^i - 1}{2^k - 2^i} \le \sum_{j = 1}^{k-1} \frac{1}{2^{j} - 1}\le \sum_{j = 1}^{\infty} \frac{1}{2^{j} - 1} = c.
\end{align}
In \cite[Theorem 2]{Heidarzadeh2019}, Heidarzadeh \emph{et al.} showed that by constructing random linear codes for which column vectors of the parity check matrix for erased symbols are linearly independent, the average blocklength of the corresponding $(l, \N, 2^k, 0)$ VLSF code is given by $\frac{k+c}{C}$. This indicates that Heidarzadeh's random linear coding scheme performs as good as Devassy's RLFC scheme for sufficiently large message length $k$.

The state-of-the-art converse bound for $(l, \N, 2^k, 0)$ VLSF codes over a BEC is also obtained by Devassy \emph{et al.} using the method of binary sequential hypothesis testing.
\begin{theorem}[Corollary 6, \cite{Devassy2016}]\label{theorem: BEC converse}
  The minimum average blocklength $l_f^*(M, 0)$ of an $(l, \N, M, 0)$ VLSF code over a BEC$(p)$ is given by
  \begin{align}
    l^*_f(M, 0) = \frac{\lfloor\log_2M\rfloor + 2\left(1 - 2^{\lfloor \log_2 M\rfloor - \log_2M}\right)}{C}.
  \end{align}
\end{theorem}
Note that when $M$ is a power of $2$, Theorem \ref{theorem: BEC converse} implies that the converse bound on maximal achievable rate is simply the capacity of the BEC.

\section{Achievable Rates of VLSF codes over BECs}\label{sec: main results}

In this section, we present a new coding scheme for a BEC called the ST-RLFC, a new achievability bound for zero-error VLSF codes of infinite decoding times, and the comparison with Devassy \emph{et al.}'s result. Finally, we present a new achievability bound for VLSF code with finite decoding times and a comparison of achievability bounds for various numbers of decoding times.

\subsection{ST-RLFC Scheme}

Consider transmitting a $k$-bit message
\begin{align}
  \bm{b} = (b_1, b_2, \dots, b_k)\in\{0, 1\}^k. \label{eq: k_bit msg}
\end{align}
Let us define the set of nonzero base vectors in $\{0, 1\}^k$ by
\begin{align}
  \G_k \triangleq \{\bm{v}\in\{0, 1\}^k:\bm{v}\ne\bm{0} \}.
\end{align}
Using ST-RLFC scheme, the channel input at time instant $n\in\N_+$ for message $\bm{b}$ is given by
\begin{align}
  X_n = \begin{cases}
    b_n, & \text{if } 1\le n\le k\\
    \bigoplus_{i=1}^k g_{n,i}b_i & \text{if } n > k,
  \end{cases} \label{eq: ST-RLFC encoder}
\end{align}
where $\oplus$ denotes bit-wise exclusive-or (XOR) operator, and $\bmg_n = (g_{n,1}, g_{n,2}, \dots, g_{n,k})^\top\in\G_k$ is generated at time instant $n$ according to a uniformly distributed random variable $\tilde{U}\in \G_k$. Note that the encoder and decoder share the same common random variable $\tilde{U}$ at time instant $n > k$ so that the decoder can produce the same $\bmg_n$ at time $n$. For $1\le n \le k$, both the encoder and decoder simply use the natural base vector $\bm{e}_n\in\R^{k\times 1}$. For all $\bm{b}\in\{0,1\}^k$, the procedure \eqref{eq: ST-RLFC encoder} specifies the common codebook before the start of transmission, i.e., the common randomness random variable $U$ in Definition \ref{def: VLSF code}.

Let $Y_n$ be the received symbol after transmitting $X_n$ over a BEC$(p)$, $p\in[0, 1)$. We consider the \emph{rank decoder} \cite{Devassy2016} that keeps track of the rank of generator matrix $G$ associated with received symbols $Y^n = (Y_1, Y_2, \dots, Y_n)$. Let $G(n)$ denote the $n$th column of $G$. If $Y_n = ?$, $G(n) = \bm{0}$; otherwise, $G(n) = \bmg_n$. Define the stopping time
\begin{align}
  \tau \triangleq \inf\{n\in\N : \text{$G(1:n)$ has rank $k$}\}, \label{eq: ST-RLFC stopping time}
\end{align}
where $G(i:j)$ denotes the matrix formed by column vectors from time instants $i$ to $j$, $1\le i\le j$. Thus, the rank decoder stops transmission at time instant $\tau$ and reproduces the $k$-bit message $\bm{b}$ using $Y^\tau$ and the inverse of $G(1:\tau)$ (namely, by solving $k$ message bits from $k$ linearly independent equations). Clearly, the error probability associated with the ST-RLFC scheme is zero.

\subsection{Achievability of Zero-Error VLSF Codes}

The ST-RLFC scheme implies the following achievability bound for $(l, \N, 2^k, 0)$ VLSF codes operating over a BEC.
\begin{theorem}\label{theorem: new BEC achiev bound}
  For a given integer $k\ge 1$, there exists an $(l, \N, 2^k, 0)$ VLSF code for BEC$(p)$, $p\in[0, 1)$, with
  \begin{align}
    l \le k + \frac{1}{C}\sum_{i=0}^{k-1}\frac{2^k-1}{2^k-2^i}F(i; k, 1-p). \label{eq: new BEC bound}
  \end{align}
  where $C = 1 - p$ and
  \begin{align}
    F(i; k, 1-p) \triangleq \sum_{j=0}^{i}\binom{k}{j}(1-p)^{j}p^{k-j} \label{eq: th13_CDF}
  \end{align}
  denotes the CDF evaluated at $i$, $0\le i\le k$, of a binomial distribution with $k$ trials and success probability $1-p$.
\end{theorem}

\begin{IEEEproof}
  See Section \ref{subsec: proof of main theorem}.
\end{IEEEproof}

For non-vanishing error probability $\epsilon > 0$, using Polyanskiy's early termination scheme in \cite[Section III-D]{Polyanskiy2011} by stopping the zero-error VLSF code at $\tau = 0$ with probability $\epsilon$, the corresponding achievability bound can be readily obtained by multiplying the right-hand side (RHS) of \eqref{eq: new BEC bound} by a factor $(1 - \epsilon)$.

We remark that the new achievability bound \eqref{eq: new BEC bound} is tighter than Devassy's bound \eqref{eq: Devassy bound} and two bounds are equal if $p = 1$ or $k = 1$. This is stated in the following corollary.
\begin{corollary}\label{corollary: tighter bound}
  For a given $k\in\N_+$ and BEC$(p)$, $p\in[0, 1]$, it holds that
  \begin{align}
    kC + \sum_{i=0}^{k-1}\frac{2^k-1}{2^k-2^i}F(i; k, 1-p) \le k + \sum_{i=1}^{k-1}\frac{2^i-1}{2^k-2^i},
  \end{align}
  where $C = 1 - p$ and $F(i; k, 1-p)$ is given by \eqref{eq: th13_CDF}. Equality holds if $p = 1$ or $k = 1$.
\end{corollary}

\begin{IEEEproof}
  Fix $k\in\N_+$ and $p\in[0, 1]$. First, note that
  \begin{align}
    k + \sum_{i=1}^{k-1}\frac{2^i-1}{2^k-2^i} = \sum_{i=0}^{k-1}\frac{2^k-1}{2^k - 2^i}.
  \end{align}
  Hence,
  \begin{align}
    &\sum_{i=0}^{k-1}\frac{2^k-1}{2^k - 2^i} - \sum_{i=0}^{k-1}\frac{2^k-1}{2^k-2^i}F(i; k, 1-p) - k(1 - p) \notag\\
    &= \sum_{i=0}^{k-1}\frac{2^k-1}{2^k-2^i}F^c(i; k, 1-p) - k(1 - p) \label{eq: cor2_eq0} \\
    &\ge \sum_{i=0}^{k-1}F^c(i; k, 1-p) - k(1 - p) \label{eq: cor2_eq1} \\
    &= 0, \notag
  \end{align}
  where in \eqref{eq: cor2_eq1}, $F^c(\cdot) \triangleq 1 - F(\cdot)$ denotes the tail probability, and the sum of tail probability equals the expectation $k(1-p)$. Note that \eqref{eq: cor2_eq0} equals $0$ if $p = 1$ or $k = 1$. This completes the proof of Corollary \ref{corollary: tighter bound}.
\end{IEEEproof}

\begin{figure}[t]
\centering
\includegraphics[width=0.48\textwidth]{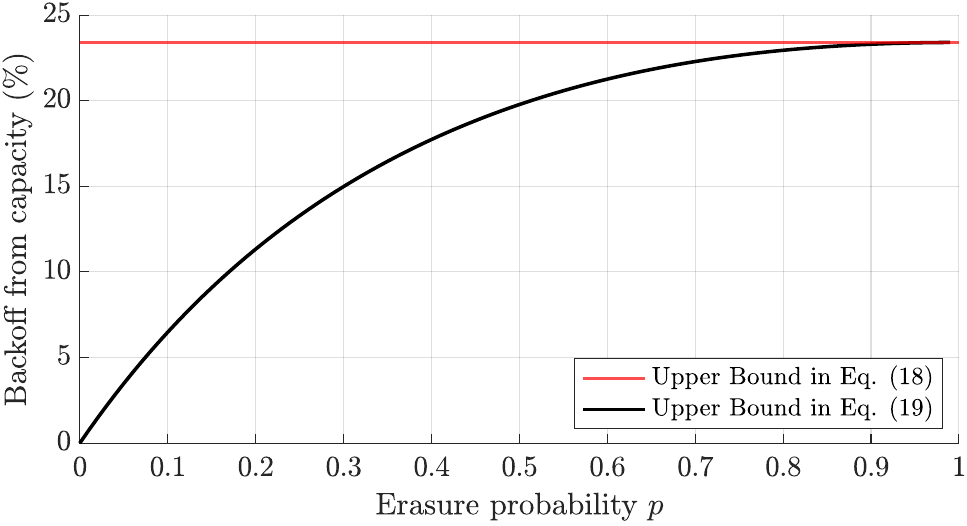}
\caption{Percentage of backoff from the capacity of BEC for $k = 3$. The red curve corresponds to a backoff percentage $23.4\%$.}
\label{fig: backoff from capacity}
\end{figure}

\begin{figure}[t]
\centering
\includegraphics[width=0.48\textwidth]{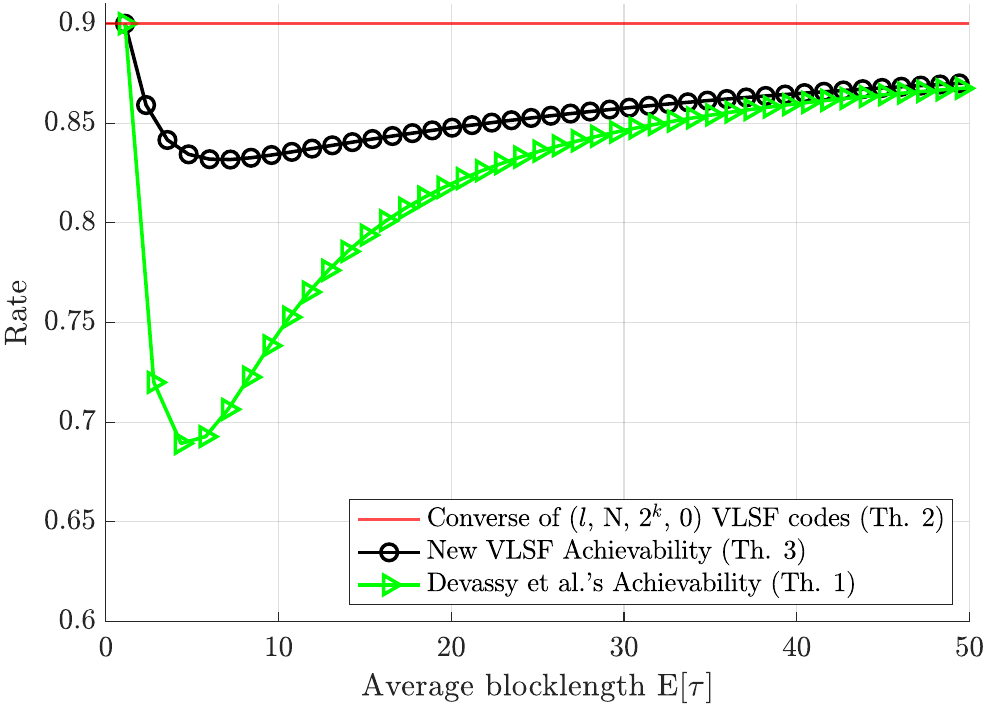}
\caption{Rate vs. average blocklength $\E[\tau]$ for $(l, \N, 2^k, 0)$ VLSF codes operated over the BEC$(0.1)$. The markers correspond to integer message lengths.}
\label{fig: rate_vs_blocklength}
\end{figure}

For BEC$(0)$ and $k\ge 2$, our new bound \eqref{eq: new BEC bound} reduces to $k$, whereas Devassy \emph{et al}'s bound \eqref{eq: Devassy bound} is strictly larger than $k$ and approaches $k + c$ for sufficiently large $k$, where $c$ denotes the Erd\"os-Borwein constant. Moreover, \eqref{eq: Devassy bound} also implies an upper bound independent of $p$ on the backoff percentage to capacity,
\begin{align}
  1 - \frac{R}{C}\le 1 - \frac{k}{k + \sum_{i=1}^{k-1}\frac{2^i-1}{2^k-2^i} }. \label{eq: old backoff percentage}
\end{align}
Devassy \emph{et al.} \cite{Devassy2016} reported that this upper bound attains its maximum $23.4\%$ at $k = 3$ and that the maximum is independent of erasure probability, thus raising the question whether this backoff percentage is fundamental. In contrast, our result in \eqref{eq: new BEC bound} implies a refined upper bound on backoff percentage that is dependent on $p$,
\begin{align}
  1 - \frac{R}{C} &\le 1 - \frac{k}{\sum_{i=0}^{k-1}\frac{2^k-1}{2^k-2^i}F(i; k, 1-p) + k(1-p)}. \label{eq: new backoff percentage}
\end{align}
Fig. \ref{fig: backoff from capacity} shows the comparison of these two upper bounds at $k  = 3$. We see that for $k = 3$, the upper bound in \eqref{eq: new backoff percentage} is a strictly increasing function of $p$. As $p\to 0$, this upper bound converges to $0$, which closes the backoff from capacity at $k = 3$. As $p \to 1$, the upper bound in \eqref{eq: new backoff percentage} converges to the backoff percentage in \eqref{eq: old backoff percentage}, as shown in Corollary \ref{corollary: tighter bound}.

\begin{figure}[t]
\centering
\includegraphics[width=0.48\textwidth]{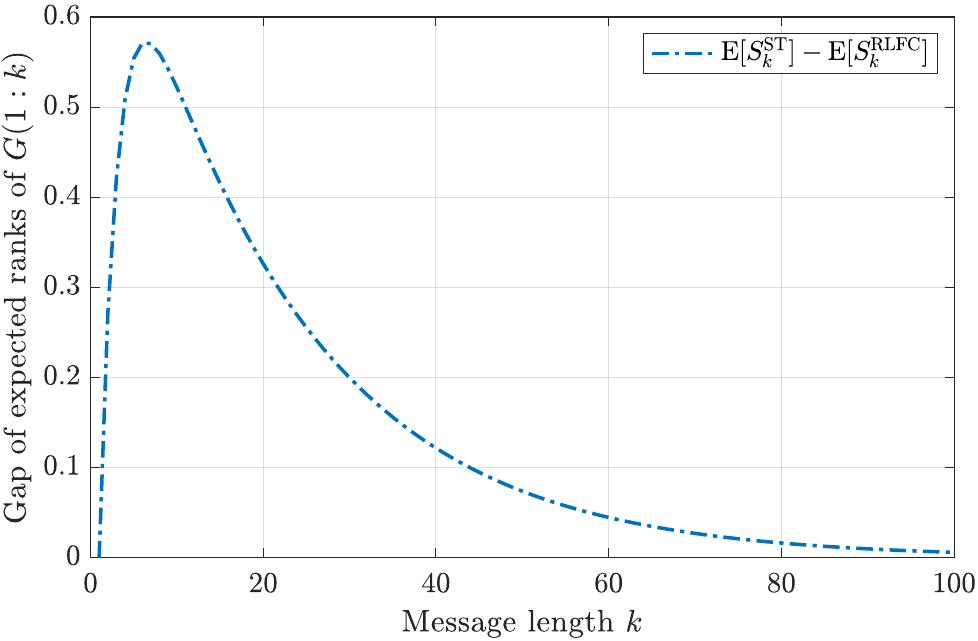}
\caption{$\E[S^{\text{ST}}_k] - \E[S^{\text{RLFC}}_k]$ vs. $k$ for BEC$(0.1)$, where $k$ ranges from $1$ to $100$.}
\label{fig: gap_expected_rank_vs_k}
\end{figure}

Fig. \ref{fig: rate_vs_blocklength} shows the comparison between the new achievability bound (Theorem \ref{theorem: new BEC achiev bound}) and Devassy \emph{et al.}'s bounds (Theorems \ref{theorem: BEC achievability} and \ref{theorem: BEC converse}) for BEC$(0.1)$. As can be seen, for a small message length, the new achievability bound is closer to capacity than Devassy \emph{et al.}'s bound. This is because when $k$ is small, systematic transmission of the uncoded message symbol is more likely to increase rank than transmitting a fountain code symbol. However, as $k$ gets larger, the advantage of systematic transmission over RLFC gradually diminishes. To see this more clearly, let random variables $S^{\text{ST}}_k$ and $S^{\text{RLFC}}_k$ denote the rank of generator matrix $G(1:k)$ for ST and RLFC, respectively. We use $\E[S^{\text{ST}}_k] - \E[S^{\text{RLFC}}_k]$ as the metric to measure the difference of rank increase rate over time range $1$ to $k$. Note that $\E[S^{\text{ST}}_k] = k(1-p)$ since the rank distribution at time instant $k$ is binomial. $\E[S^{\text{RLFC}}_k]$ can be numerically computed using the one-step transfer matrix $P$ in \eqref{eq: ap2_one_step_transfer_matrix} and initial distribution $[1, \bm{0}^\top]\in\R^{1\times (k+1)}$. Fig. \ref{fig: gap_expected_rank_vs_k} shows $\E[S^{\text{ST}}_k] - \E[S^{\text{RLFC}}_k]$ as a function of message length $k$ for BEC$(0.1)$. We see that this gap constantly remains nonnegative, implying that the rank increase rate for ST is always faster than that for RLFC. For sufficiently large $k$, this gap becomes small, indicating the diminishing advantage of ST over RLFC.

\subsection{Achievability of VLSF Codes with Finite Decoding Times}

The ST-RLFC scheme also facilitates an $(l, n_1^m, 2^k, \epsilon)$ VLSF code for a BEC. This code is constructed by using the same ST-RLFC scheme in \eqref{eq: ST-RLFC encoder} but a rank decoder that only considers a finite set of decoding times. Specifically, fix $n_1^m\in\N_+$ satisfying $n_1 < n_2 < \cdots < n_m$. For a given $k\in\N_+$ and $\epsilon\in(0, 1)$, the rank decoder still shares the same common randomness with the encoder in selecting the base vector $\bmg_n$, except that it now adopts the following stopping time:
\begin{align}
  \tau^* \triangleq \inf\{n\in\{n_i\}_{i=1}^m: G(1:n) \text{ has rank $k$ or $n = n_m$} \}. \label{eq: new stopping time}
\end{align}
If $\tau \le n_m$ and $G(1:\tau)$ is full rank, the rank decoder reproduces the transmitted message using $Y^\tau$ and the inverse of $G(1:\tau)$. If $\tau = n_m$ and $G(1:n_m)$ is rank deficient, then the rank decoder outputs an arbitrary message.

The ST-RLFC scheme and the modified rank decoder imply the following achievability bound for an $(l, n_1^m, 2^k, \epsilon)$ VLSF code.
\begin{theorem}\label{theorem: new nonasymptotic achiev bound for BEC}
  Fix $n_1^m\in\N_+^m$ satisfying $n_1 < n_2 < \cdots < n_m$. For any positive integer $k\in\N_+$ and $\epsilon\in(0, 1)$, there exists an $(l, n_1^m, 2^k, \epsilon)$ VLSF code for the BEC$(p)$ with
    \begin{align}
      l &\le n_m - \sum_{i=1}^{m-1}(n_{i+1} - n_i)\Prob[S_{n_i} = k], \label{eq: nonasymptotic bound on l} \\
      \epsilon &\le 1 - \Prob[S_{n_m} = k],  \label{eq: nonasymptotic bound on epsilon}
    \end{align}
  where the random variable $S_n$ denote the rank of the generator matrix $G(1:n)$ observed by the rank decoder. Specifically, $\Prob[S_n = k]$ is given by
  \begin{align}
    \Prob[S_n = k] = \begin{cases}
      0, & \text{if } n < k \\
      1 - \bm{\alpha}^\top T^{n-k}\bm{1}, &\text{if } n\ge k,
    \end{cases} \label{eq: expression for S_n}
  \end{align}
where $\bm{\alpha} = [\alpha_1,\alpha_2,\dots, \alpha_k]^\top\in\R^{k\times 1}$ with $\alpha_i = F(i; k, 1-p)$, $0\le i\le k-1$, and $F(i;k, 1-p)$ is given by \eqref{eq: th13_CDF}, $T\in\R^{k\times k}$ with entries given by
  \begin{align}
    T_{i, i} &= p + \frac{(1-p)(2^{i-1} - 1)}{2^k - 1}, \\
  T_{i,i+1} &= \frac{(1-p)(2^k - 2^{i-1})}{2^k - 1}, \\
  T_{i,j} &= 0, \text{ for } j\ne i \text{ and } j\ne i+1.
  \end{align}
\end{theorem}

\begin{IEEEproof}
  See Section \ref{subsec: proof of second result}.
\end{IEEEproof}

Theorem \ref{theorem: new nonasymptotic achiev bound for BEC} facilitates an integer program that can be used to compute the achievability bound on rate for all zero-error VLSF codes of message size $M = 2^k$ and $m$ decoding times. Define
\begin{align}
  N(n_1^m) &\triangleq n_m - \sum_{i=1}^{m-1}(n_{i+1} - n_{i})\Prob[S_{n_i} = k].
\end{align}
For a given number of decoding times $m\in\N_+$, message length $k\in\N_+$, and a target error probability $\delta\in(0,1)$, 
\begin{align}
  \begin{split}
    \min_{n_1^m}&\quad N(n_1^m) \\
    \st&\quad 1 - \Prob[S_{n_m} = k]\le \delta
  \end{split} \label{eq: ST-RLFC integer program}
\end{align}
Assume $N(a_1^m)$ is the minimum value after solving \eqref{eq: ST-RLFC integer program}, where $a_1^m$ is the minimizer. Then the achievability bound on rate for a VLSF code of message size $2^k$, target error probability $\delta$, and $m$ decoding times is given by $\frac{k}{N(a_1^m)}$. For reasonably small values of $m$, one can use brute-force method to obtain $a_1^m$ and $N(a_1^m)$.

\begin{figure}[t]
\centering
\includegraphics[width=0.48\textwidth]{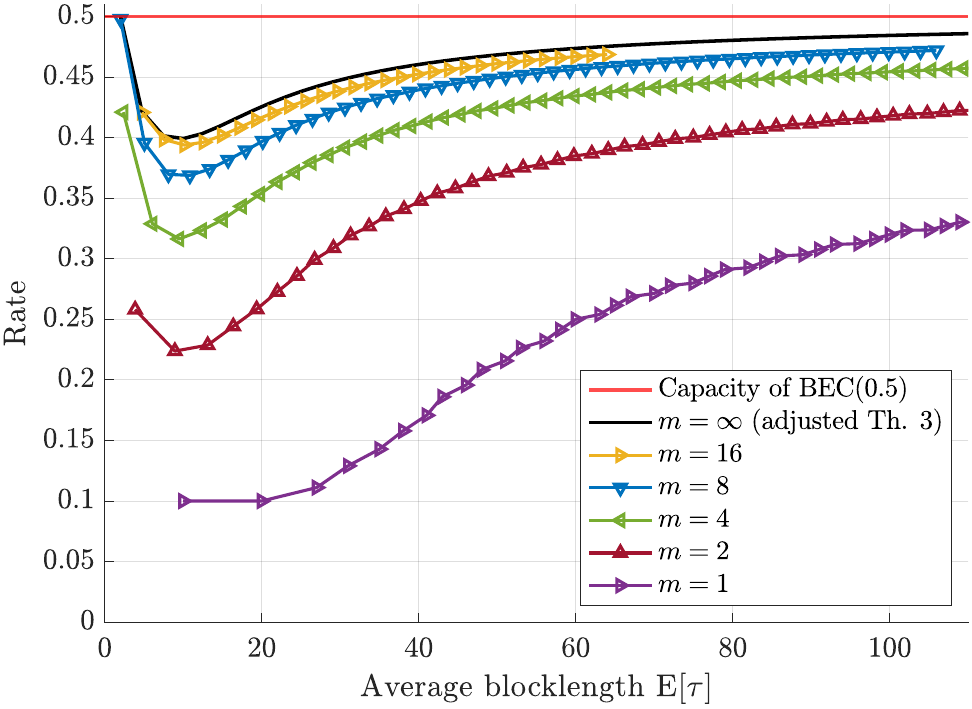}
\caption{Rate vs. average blocklength $\E[\tau]$ for $(l, n_1^m, 2^k, \delta)$ VLSF codes operated over the BEC$(0.5)$, where $\delta = 10^{-3}$. The marker corresponds to integer message length $k$. In this figure, $k\ge1$ for $m = 1,2,4,8$ and $k\ge2$ for $m = 16$. Legend ``adjusted Th. 3'' means that the upper bound on $l$ is the RHS of \eqref{eq: new BEC bound} multiplied by $(1 - \delta)$.}
\label{fig: rate_vs_blocklength_finite_m}
\end{figure}

For BEC$(0.5)$, Fig. \ref{fig: rate_vs_blocklength_finite_m} shows achievability bounds for $(l, n_1^m, 2^k, \delta)$ VLSF codes, where $m\in\{1,2,4,8, 16\}$ and target error probability $\delta = 10^{-3}$. The adjusted achievability bound boosted by $\frac{1}{1-\delta}$ for $m = \infty$ using Theorem \ref{theorem: new BEC achiev bound} and Polyanskiy's early termination scheme is also shown. We see that when $m$ is small, increasing $m$ can dramatically improve the achievable rate. However, when $m = 16$, the achievable rate closely approaches that for $m = \infty$. We remark that similar effect on achievable rate by the varying number of decoding times has also been observed in several previous works, e.g., \cite{Heidarzadeh2019,Yavas2021,Yang_ISIT2022}.

\section{Proofs}\label{sec: proofs}

In this section, we prove our main results.

\subsection{Proof of Theorem \ref{theorem: new BEC achiev bound}} \label{subsec: proof of main theorem}

Let random variable $S_n$ denote the rank of generator matrix $G(1:n)$. According to the ST-RLFC scheme, the probability mass function (PMF) of $S_k$ at time $k$ is given by
\begin{align}
  \Prob[S_k = r] = \binom{k}{r}(1-p)^rp^{k-r},\quad 0\le r\le k. \label{eq: ap2_eq1}
\end{align}
For $n\ge k$, due to the BEC$(p)$ and our RLFC scheme, $S_{n+1} = S_n = r$ occurs if $Y_{n+1} = ?$ or if $Y_{n+1}\ne ?$ and $\bmg_{n+1}$ is a linear combination of previous $r$ independent base vectors. Otherwise, $S_{n+1} = r + 1$. Hence, the behavior of $S_n$, $n\ge k$, is characterized by the following discrete-time homogeneous Markov chain with $k+1$ states.
\begin{align}
  &\Prob[S_{n+1} = r|S_n = r] = p + \frac{(1-p)(2^r - 1)}{2^k - 1}, \\
  & \Prob[S_{n+1} = r+1|S_n = r] = \frac{(1-p)(2^k - 2^r)}{2^k - 1}, \label{eq: rank increase prob}
\end{align}
where $0\le r\le k-1$, and $\Prob[S_{n+1} = k|S_n = k] = 1$. Note that this Markov chain has a single absorbing state $S_n = k$. The time to absorption for this Markov chain follows a discrete phase-type distribution \cite[Chapter 2]{Neuts_book}. More specifically, the one-step transfer matrix $P\in\R^{(k+1)\times (k+1)}$ of this Markov chain can be written as
\begin{align}
  P = \begin{bmatrix}
    T & \bm{t} \\
    \bm{0}^\top & 1
  \end{bmatrix}, \label{eq: ap2_one_step_transfer_matrix}
\end{align}
where the entries of $T\in\R^{k\times k}$ are given by
\begin{align}
  T_{i, i} &= p + \frac{(1-p)(2^{i-1} - 1)}{2^k - 1}, \\
  T_{i,i+1} &= \frac{(1-p)(2^k - 2^{i-1})}{2^k - 1},
\end{align}
and $T_{i,j} = 0$ for any other pair $(i, j)$, $1\le i,j\le k$. Since $P$ is a stochastic matrix, it follows that
\begin{align}
  \bm{t} = (I -T)\bm{1}.
\end{align}
The initial probability distribution is given by $[\bm{\alpha}^\top, \alpha_{k}]$, where
\begin{align}
  \bm{\alpha}^\top \triangleq \begin{bmatrix}
    \Prob[S_k=0] & \Prob[S_k=1] & \cdots & \Prob[S_k=k-1]
  \end{bmatrix}, \label{eq: initial distribution}
\end{align}
with $\Prob[S_k = r]$ given by \eqref{eq: ap2_eq1}, and $\alpha_k = 1 - \bm{\alpha}^\top\bm{1}$. Let random variable $X\in\N$ denote the time to absorbing state $k$ with initial distribution $[\bm{\alpha}^\top, \alpha_{k}]$. Hence, it follows that $X$ has PMF
\begin{align}
  \Prob[X = n] = \bm{\alpha}^\top T^{n-1}\bm{t},\quad n\in\N_+,
\end{align}
and $\Prob[X = 0] = \alpha_{k}$. Define the generating function of $X$ by
\begin{align}
  H_X(z)&\triangleq \E[z^X] = \sum_{n=0}^\infty z^n\Prob[X = n] \notag\\
    &= \alpha_k + \sum_{n=1}^{\infty}z^n\bm{\alpha}^\top T^{n-1}\bm{t} \notag\\
    &= \alpha_k + z\bm{\alpha}^\top\Paren{\sum_{n=0}^\infty (zT)^n }\bm{t} \label{eq: ap2_eq2} \\
    &= \alpha_k + z\bm{\alpha}^\top(I - zT)^{-1}(I - T)\bm{1},
\end{align}
where in \eqref{eq: ap2_eq2}, we have used $\sum_{n=0}^\infty A^n = (I - A)^{-1}$ whenever $|\lambda_i| < 1$ for all $i\in [k]$, where $\{\lambda_i\}_{i=1}^k$ denotes the eigenvalues of a square matrix $A\in\R^{k\times k}$. Hence, the expected time to absorbing state $k$ is given by
\begin{align}
  \E[X] &= \frac{\diff H_X(z)}{\diff z}\Big|_{z = 1} = \bm{\alpha}^\top(I - T)^{-1}\bm{1}.
\end{align}
Therefore, the expected stopping time $\E[\tau]$, with $\tau$ defined in \eqref{eq: ST-RLFC stopping time}, is given by
\begin{align}
  \E[\tau] &= k + \E[X] \notag\\
    &= k + \bm{\alpha}^\top(I - T)^{-1}\bm{1}  \label{eq: ap2_eq3}
\end{align}
Note that 
\begin{align}
  I - T &= (1 - p)\diag\Paren{1, \frac{2^k-2^1}{2^k-1},\frac{2^k-2^2}{2^k-1},\cdots, \frac{2^k-2^{k-1}}{2^k-1} } \notag\\
    &\phantom{==}\cdot\begin{bmatrix}
      1 & -1 & 0 & \cdots & 0 \\
      0 & 1 & -1 & \cdots & 0 \\
      0 & 0 & 1 & \cdots & 0 \\
      \vdots & \vdots & \vdots & \ddots & \vdots \\
      0 & 0 & 0 & \cdots & 1
    \end{bmatrix}
\end{align}
Hence, 
\begin{align}
  &(I - T)^{-1} = (1-p)^{-1}\begin{bmatrix}
      1 & 1 & 1 & \cdots & 1 \\
      0 & 1 & 1 & \cdots & 1 \\
      0 & 0 & 1 & \cdots & 1 \\
      \vdots & \vdots & \vdots & \ddots & \vdots \\
      0 & 0 & 0 & \cdots & 1
    \end{bmatrix}\notag\\
    &\phantom{==}\cdot\diag\Paren{1, \frac{2^k-1}{2^k-2^1}, \frac{2^k-1}{2^k-2^2},\cdots \frac{2^k-1}{2^k - 2^{k-1}} } \notag\\
  &= (1-p)^{-1}\begin{bmatrix}
      1 & \frac{2^k-1}{2^k-2^1} & \frac{2^k-1}{2^k-2^2} & \cdots & \frac{2^k-1}{2^k - 2^{k-1}} \\
      0 & \frac{2^k-1}{2^k-2^1} & \frac{2^k-1}{2^k-2^2} & \cdots & \frac{2^k-1}{2^k - 2^{k-1}} \\
      0 & 0 & \frac{2^k-1}{2^k-2^2}  & \cdots & \frac{2^k-1}{2^k - 2^{k-1}} \\
      \vdots & \vdots & \vdots & \ddots & \vdots \\
      0 & 0 & 0 & \cdots & \frac{2^k-1}{2^k - 2^{k-1}}
    \end{bmatrix}. \label{eq: inverse matrix}
\end{align}
Substituting \eqref{eq: initial distribution} and \eqref{eq: inverse matrix} into \eqref{eq: ap2_eq3}, we finally obtain
\begin{align}
  \E[\tau] &= k + (1-p)^{-1}\sum_{i=0}^{k-1}\frac{2^k-1}{2^k - 2^i}\sum_{j=0}^i\Prob[S_k = j] \\
    &= k + \frac{1}{C}\sum_{i=0}^{k-1}\frac{2^k-1}{2^k - 2^i}F(i; k, 1-p), \label{eq: ap2_eq4}
\end{align}
where $C = 1 - p$ and $F(i;k,1-p) \triangleq \sum_{j=0}^i\Prob[S_k = j]$ denotes the CDF evaluated at $i$ of a binomial distribution with $k$ trials and success probability $1 - p$. Since \eqref{eq: ap2_eq4} is the expected stopping time for an ensemble of zero-error VLSF codes, there exists an $(l, \N, 2^k, 0)$ VLSF code with
\begin{align}
  l \le k + \frac{1}{C}\sum_{i=0}^{k-1}\frac{2^k-1}{2^k - 2^i}F(i; k, 1-p).
\end{align}
This concludes the proof of Theorem \ref{theorem: new BEC achiev bound}.

\subsection{Proof of Theorem \ref{theorem: new nonasymptotic achiev bound for BEC}} \label{subsec: proof of second result}

The proof essentially builds upon that of Theorem \ref{theorem: new BEC achiev bound} with the distinction that the rank decoder adopts a new stopping time given by \eqref{eq: new stopping time}.

Let $S_n$ denote the rank of the generator matrix $G(1:n)$ observed at the rank decoder. The expected stopping time $\E[\tau^*]$ is written as
\begin{align}
  \E[\tau^*] &= \sum_{n=0}^\infty\Prob[\tau^* > n] \notag\\
    &= n_1 + \sum_{i=1}^{m-1}(n_{i+1} - n_i)\Prob[\tau^* > n_i] \\
    &= n_1 + \sum_{i=1}^{m-1}(n_{i+1} - n_i)\Prob[S_{n_i} < k] \\
    &= n_m - \sum_{i=1}^{m-1}(n_{i+1} - n_{i})\Prob[S_{n_i} = k],
\end{align}
thus proving the upper bound in \eqref{eq: nonasymptotic bound on l}.

Note that at finite blocklength, the error only occurs when the rank of generator matrix $G(1:n_m)$ is still less than $k$. Hence,
\begin{align}
  \epsilon &\le \Prob[S_{n_m} < k] \\
    &= 1 - \Prob[S_{n_m} = k],
\end{align}
which is equal to the upper bound in \eqref{eq: nonasymptotic bound on epsilon}. 

At time $n < k$, due to the systematic transmission, $\Prob[S_n = k] = 0$. At time $n\ge k$, as discussed in Section \ref{subsec: proof of main theorem}, the behavior of $S_n$ is characterized by a discrete-time homogeneous Markov chain with $k+1$ states whose one-step transfer matrix is given by \eqref{eq: ap2_one_step_transfer_matrix}, and whose initial probability distribution is $[\bm{\alpha}^\top, \alpha_k]$, where $\bm{\alpha}^\top$ is given by \eqref{eq: initial distribution}. Hence, for $n \ge k$,
\begin{align}
  \Prob[S_n = k] &= 1 - \Prob[S_n < k] \\
    &= 1 - \bm{\alpha}^\top T^{n-k}\bm{1}.
\end{align}
This completes the proof of Theorem \ref{theorem: new nonasymptotic achiev bound for BEC}.

\section{Conclusion}\label{sec: conclusion}

Using the ST-RLFC scheme and the rank decoder, we have shown an improved achievability bound for zero-error VLSF codes of message size $M = 2^k$. The improvement leverages the fact that when $k$ is small, initially transmitting systematic message symbols is more likely to increase the rank of the generator matrix than transmitting fountain code symbols. However, as demonstrated in Fig. \ref{fig: rate_vs_blocklength}, there is still a significant gap between the converse and the achievability bounds. It remains to be seen how to close this gap. In addition, the extension of Theorem \ref{theorem: new BEC achiev bound} to arbitrary message size $M$ still remains open.

The ST-RLFC scheme combined with a modified rank decoder facilitates a VLSF code of finite decoding times and bounded error probability. Fig. \ref{fig: rate_vs_blocklength_finite_m} shows that when $m$ is small, a slight increase in $m$ can dramatically improve the achievable rate. On the other hand, when $m$ is moderately large (for instance, $m = 16$ for BEC$(0.5)$ shown in Fig. \ref{fig: rate_vs_blocklength_finite_m}), the achievable rate closely approaches that for $m = \infty$. However, a proof that shows this trend still remains elusive.






\ifCLASSOPTIONcaptionsoff
  \newpage
\fi

\bibliographystyle{IEEEtran}
\bibliography{IEEEabrv,reference}

\end{document}